# Magnetoelectric effect in van der Waals magnets

Kai-Xuan Zhang[1,2,3]★$, Giung Park[1,2,3], Youjin Lee[1,2,3], Beom Hyun Kim[4], and Je-Geun Park[1,2,3]★

[1]*Department of Physics and Astronomy, Seoul National University, Seoul 08826, South Korea*

[2]*Center for Quantum Materials, Department of Physics and Astronomy, Seoul National University, Seoul 08826, South Korea*

[3]*Institute of Applied Physics, Seoul National University, Seoul 08826, South Korea*

[4]*Center for Theoretical Physics of Complex Systems, Institute for Basic Science, Daejeon, 34126, Korea*

$*Present address: Department of Physics, Washington University in St. Louis, St. Louis, Missouri 63130, USA*

★ Corresponding authors: Kai-Xuan Zhang (email: kxzhang.research@gmail.com) and Je-Geun Park (email: jgpark10@snu.ac.kr).

**Abstract:** The magnetoelectric (ME) effect is a fundamental concept in modern condensed matter physics and represents the electrical control of magnetic polarisations or vice versa. Two-dimensional (2D) van-der-Waals (vdW) magnets have emerged as a new class of materials and exhibit novel ME effects with diverse manifestations. This review emphasizes some important recent discoveries unique to vdW magnets: multiferroicity on two dimensions, spin-charge correlation, atomic ME effect and current-induced intrinsic spin-orbit torque, and electrical gating control and magnetic control of their electronic properties. We also highlight the promising route of utilizing quantum magnetic hetero- or homo-structures to engineer the ME effect and corresponding spintronic and optoelectronic device applications. Due to the intrinsic two-dimensionality, vdW magnets with those ME effects are expected to form a new, exciting research direction.

# 1. Introduction

The magnetoelectric (ME) effect was discovered a century ago and entered a renaissance with the advent of multiferroic materials[1]. ME effect manifests as the electrical control of magnetic polarisations or, reversely, the magnetic regulation of electrical polarisations. Such a feature intrinsically reflects the coupling of spin and charge degrees of freedom or more. As such, the ME effect is significant in both the fundamental physics of correlation and device applications, including magnetic sensors, logic and spintronic devices, and tunable microwave filters.

Before going straight into the ME effect found in vdW magnets, it would benefit the public if we discuss the more traditional multiferroic physics often discussed in the context of 3D magnets. A more usual definition of multiferroic is based on the linear ME effect that couples two otherwise independent degrees of freedom, spin and charge. This coupling, in principle, allows the control of charge by spin or vice versa, of which potential applications have been the source of much enthusiasm in the field. A dominant mechanism behind this interesting physics is the spin current model, where a noncollinear spiral magnetic structure produces an electric dipole via inverse Dzyalloshin-Moriya interaction[2]. Another equally viable mechanism is through a spin-lattice or magnon-phonon coupling[3]. Another noteworthy point about the 3D multiferroic materials is the classification. Two popular schemes are used: type I/II multiferroic and proper/improper multiferroic. The many multiferroic compounds can be divided into two groups[4]: type I multiferroics, where the ferroelectricity and magnetism occur at different temperatures and with different mechanisms, and type II multiferroics, where a magnetic order causes ferroelectricity.

Unlike these 3D multiferroic materials[5,6], recent two-dimensional (2D) van-der-Waals (vdW) magnets[7-11] pose the new potential to develop a more colourful landscape of the old-yet-classical ME effect and inspire many new opportunities in designing/manifesting/utilizing the ME effect from diverse angles of view. Most importantly, its intrinsic two-dimensionality offers a unique possibility that would otherwise be difficult, if not impossible, using bulk materials. In this review, we deliberately take a much broader view of ME effects than typically adopted in discussing 3D multiferroic materials. The benefit of adopting this broader approach is that it will allow us to cover more diverse and novel phenomena. This way, we can highlight some fascinating opportunities with the spin-charge coupling in 2D vdW magnets.

For example, $NiI_2$ has been discovered as the first vdW multiferroic stable to very thin layers with a second harmonic generation response maintaining a free-standing bilayer[12] and even an hBN substrate-supported monolayer[13]. The ME physics of vdW magnets can be extended to other richer quantum effects such as quantum entangled magnetic exciton[14]. Another manifestation of the ME effect resides in $NiPS_3$, where $NiPS_3$ hosts an XXZ-type antiferromagnetic spin texture and, most importantly, a unique sharp magnetic exciton from the Zhang-Rice singlet to triplet transition[15]. A careful examination reveals that a small charge transfer occurs driven by a hybrid spin transition and a spin-orbit entanglement, eventually leading to an electrical dipole polarisation of roughly 0.1~0.2 $\mu C/cm^2$ per $NiS_6$ octahedron and consequent magnetic exciton[15]. This atomic ME effect is the origin of the strong spin-charge coupling, as reported earlier[16], demonstrating a unique case of how the ME effect can arise from new quantum many-body phenomena.

Another interesting aspect of the ME effect in vdW magnets is its potential application in spintronics. For example, a novel atomic ME effect also exists in a ferromagnetic metal, e.g., the topological vdW perpendicular ferromagnet $Fe_3GeTe_2$[17]. The atomic ME coefficient was theoretically investigated in terms of its symmetry and Berry curvature of the topological bands with spin-orbit coupling[17]. Surprisingly, such ME effect can be reflected in its free energy formula and greatly modify the spin-related free energy landscape, significantly lowering the switching barrier and reducing the switching coercive field correspondingly[17]. The large Berry curvatures lead to a gigantic intrinsic spin-orbit torque (SOT)[17-19] by current in $Fe_3GeTe_2$ itself, reaching around 50 Oe/(mA/μm$^2$)[17]. Such SOT value is about 100 times larger than conventional ferromagnet/heavy-metal composite systems, guaranteeing a highly efficient nonvolatile switching by current in its magnetic memory devices[20,21]. This $Fe_3GeTe_2$ case forms another case of bridging the underlying ME effect to fascinating quantum/topological properties and advanced spintronic applications.

These above scientific cases directly manifest the intrinsic ME effect at the atomic scale. More generally, the ME effect relates to any electrical manipulation of magnetic properties and vice versa. There are many interesting examples in the vdW magnet field; the most prototypical one is solid/ionic gating[22-27]. Gating can induce more electron or hole carriers in the system, shifting the Fermi level, altering the correlations and mediating the magnetic interactions in many vdW magnets[22-26]. In addition, constructing vdW hetero- or homo-structures based on vdW magnets can be another effective way to engineer the ME effect[28-30].

This review aims to introduce these scientific examples and discuss these results in the context of the ME effect in vdW magnets. In this way, we want to emphasize the new insights and opportunities that 2D vdW magnets bring to advance and enrich the ME effect with a fresh viewpoint. Moreover, we provide more outlooks and a few examples of the ME effect's revolution using quantum combinations and designs of vdW magnets. At the same time, we want to bring attention to several new concepts like quantum, topology, correlation, symmetry, etc., to leverage the atomic ME couplings.

## 2. Atomically thin multiferroic $NiI_2$

Multiferroic materials show a coexistence of magnetic and ferroelectric order. The multiferroic systems provide an efficient route for controlling magnetism through an electric field and vice versa. Controlling magnetic order via an external electric field consumes less energy and complexity than controlling via a magnetic field. The vdW magnetic systems develop new paradigms for high-density magnetic memory due to their tiny volume and effective control of the magnetic state using external stimuli. Therefore, exploiting the vdW multiferroic material will provide promising candidates for future high-density, eco-friendly memory devices.

It is noteworthy that several halogenated compounds such as $NiI_2$, $CoI_2$, $MnI_2$, and $NiBr_2$ exhibit multiferrocity in the bulk crystals[31-33]. The multiferroic order of $NiI_2$ persists down to the bilayer[12] and even the two-dimensional limit at the specific condition[13,34] in experiments.

Bulk NiI$_2$ has two successive magnetic phase transitions: the paramagnetic-to-antiferromagnetic transition at $T_{N1}$ = 76 K and the helimagnetic transition at $T_{N2}$ = 59.5 K (Fig. 1a). At the helimagnetic state, it develops a proper-screw spin helix with propagation vector Q = (0.138,0,1.457) in reciprocal lattice units. In such a noncollinear structure, the inverse Dzyaloshinskii-Moriya (DM) interaction or the spin-dependent metal-ligand hybridization induces nonzero electric polarisation. The induced electric polarisation is perpendicular to the magnetic propagation vector Q, developing in-plane ferroelectricity with six possible ferroelectric domains.

The optical second harmonic generation (SHG) measurements reveal its multiferroicity from the broken inversion symmetry in bulk and few-layer NiI$_2$. The broken centrosymmetry from the electric dipole contribution gives rise to the nonzero SHG signal. The intermediate antiferromagnetic phase shows a small six-fold anisotropy pattern reflecting the triangular lattice symmetry. On the other hand, the multiferroic phase represents a two-fold symmetry with an enhanced SHG signal. Such an enhanced SHG corresponds to the ferroelectric order parameter, considering that SHG intensity is proportional to the square of electric polarisation. The multiferroic phase transition temperature also shows a strong thickness dependence. Figure 2 shows that the transition temperature decreases as the layer number decreases. The SHG signal persists down to the bi-layer and disappears in the monolayer. Another group also demonstrated the thickness-dependent multiferroicity in few-layer NiI$_2$ on an hBN substrate grown by physical vapour deposition (PVD)[13]. The birefringence and SHG signals from the ferroelectricity are consistent with the previous results down to the bilayer: decreasing transition temperature with reduced thickness. In this case, the multiferroicity survives down to the monolayer. The discrepancy between the two studies on single-layer NiI$_2$ from different sample preparations could be attributed to the strain effect from the substrate[35]. The following STM study also shows that single-layer NiI$_2$ on top of highly oriented pyrolytic graphite (HOPG) provides additional proof of the two-dimensional multiferroicity[34].

The multiferroic state further induces magnetic exciton in NiI$_2$[14]. Optical absorption spectra reveal an exciton peak at 1.384 eV and a two-magnon sideband peak at 1.396 eV (Fig. 3). The exciton peak appears only below $T_{N2}$, implying a connection to the multiferroic phase. NiI$_2$ forms multiferroic domains in six directions, and these possible domains determine the angular dependence of exciton. Such an angular dependency of the exciton at 1.384 eV provides strong evidence of its association with the ferroelectric spiral phase. Combined with X-ray absorption spectra (XAS) and the resonant inelastic X-ray spectroscopy (RIXS), the mechanism of creation of magnetic excitons was studied. The magnetic excitons are generated from the Zhang-Rice triplet (ZRT) to the singlet (ZRS) transition. XAS reveals a relatively small charge-transfer energy of 1.498 eV, slightly larger than NiPS$_3$ of 0.95 eV[15] but significantly smaller than NiO of 4.6 eV[36]. Such a small charge-transfer energy is a prerequisite for the ZRT state. The RIXS and configuration interaction (CI) model will validate the ZRT to ZRS transition scenario in NiI$_2$.

On the other hand, THz spectroscopy on NiI$_2$ reveals two electromagnon modes at 34 and 37 cm$^{-1}$ in the multiferroic phase at 2.5 K[37]. These energies are consistent with the energy scale of the two-magnon sideband excitation around the magnetic exciton peak. The reflection mode shows that these magnons are electric dipole active rather than magnetic dipole active. The observed electromagnon mode shines a valuable light on the origin of the very strong magnetoelectric coupling in NiI$_2$[38]. The following study employs time-resolved optical SHG and Kerr rotation microscopy to reveal the nature of the electromagnons. SHG is sensitive to the electric dipole polarisation, and Kerr rotation microscopy is sensitive to the magnetic order; therefore, analyzing these two results gives a comprehensive understanding of the electric polarisation and magnetization components. Perturbing the multiferroic state with a laser pulse triggers coherent oscillations of the electromagnon modes. Two time-resolved optical techniques, SHG and Kerr rotation, show the emergence of an π/2 phase shift between the electric polarisation and magnetization oscillations. The estimated dynamical magnetoelectric coupling strength surpasses other multiferroics, such as CuO[39]. Such a substantial enhancement originates from the synergy between the noncollinear spin texture and spin-orbit interaction on the ligands and *d-p* hybridization compared to the phonon-mediated electromagnons.

As mentioned previously, several other vdW multiferroic systems exist besides NiI$_2$. Future systematic studies will be helpful in most of these systems, such as the thickness dependence of multiferroicity as performed in NiI$_2$. It is encouraging that theoretical studies predicted that some metal halides could host multiferroicity even in the monolayer limit[40,41]. To experimentally realize multiferroicity in monolayer or few-layer metal halides, however, there are some realistic experimental challenges to overcome, at least including issues like a proper substrate for stabilizing multiferroic states and encapsulation for preventing degradations due to their air-sensitivity. Another interesting vdW multiferroic material, CuCrP$_2$S$_6$ was also studied using multiple probes such as magnetization, electrical transport, and SHG measurements, expanding the vdW multiferroic investigations[42,43]. Further exploiting the vdW multiferroic materials will stimulate new methodologies for controlling the multiferroicity in the atomic layer limit.

## 3. Atomic magnetoelectric effect in NiPS$_3$

NiPS$_3$ is a charge transfer (CT) insulator that exhibits zigzag antiferromagnetic ordering[16,44,45]. Previous studies have estimated the CT energy to be either negative[16] or within a small positive range from 0.9 to 2.5 eV[15,46,47]. The small CT energy, combined with strong *pd-σ* hybridization, causes doubly degenerate unoccupied Ni $e_g$ orbitals to mix with occupied ligand *p* orbitals from the surrounding six sulfur atoms. This leads to the spontaneous population of self-doped ligand holes[48]. The combination of a hole at the Ni site and a self-doped ligand hole forms spin-orbital entangled many-body states within the NiS$_6$ octahedron, manifesting as spin-triplet and orbital singlet states[15,47,49,50]. These CT states are reminiscent of the Zhang-Rice triplet (ZRT) states observed in high-$T_C$ cuprates[51,52]. Recent studies suggest that the ground state of NiPS$_3$ is a mixture of approximately 60% triplet states from the $d^8$ configuration and about 40% ZRT states from the $d^9L^1$ configuration, where $L$ represents the ligand hole[15,47].

In this context, we examine the self-doped ligand distribution of the ZRT states within the honeycomb lattice. When ZRT states form in each NiS$_6$ octahedron, two types of orthonormal S 3*p* holes can populate the sulfur sites, forming spin-triplet states with neighbouring Ni 3*d* orbitals. Under zigzag magnetic ordering, two edge-shared sulfur sites emerge based on the spin configurations of neighbouring Ni ions: $S_1$ sites with parallel spin alignments and $S_2$ sites with anti-parallel spin alignments. At $S_1$ sites, the two-hole orbitals share the same spin, whereas they have opposite spins at $S_2$ sites (see Fig. 4a). Charge disproportionation between these two sulfur sites induces charge-stripe modulation, which can lead to local electronic dipole polarisation. Many-body calculations have shown that exchange correlations (Hund's coupling) between self-doped holes can cause charge deviations of approximately 0.002~0.003$|e|$ between the $S_1$ and $S_2$ sites, as shown in Fig. 4c, resulting in a local dipole polarisation of around 0.1~0.2 μC/cm$^2$ per NiS$_6$ octahedron[15]. Alternatively, atomic spin-orbital coupling, combined with non-cubic hopping and spin-orbit coupling, has generated a local dipole polarisation of about 10$^{-5}$ μC/cm$^2$ in the ZRT state under zigzag magnetic ordering[50]. These induced local electronic dipoles give rise to the atomic magnetoelectric effect.

The atomic magnetoelectric effect contributes to the spin-charge coupling observed in optical excitations[16,53,54]. A prominent manifestation of this effect is the magneto-exciton behaviour in an ultra-narrow exciton peak at around 1.476 eV[15,55-60]. This exciton becomes optically bright only when the zigzag magnetic order is stabilized and exhibits two-fold intensity anisotropy under an in-plane magnetic field. The magneto-exciton behaviour can be explained by the emergent electric field induced by the reversal of local dipole polarisation during the transition from the ground ZRT state to the excited Zhang-Rice singlet state (see Fig. 4b,d). Additionally, these excitons show anisotropic Zeeman-like splitting in resonance with an in-plane magnetic field[61-63]. This splitting can also be interpreted as a magnetic exciton arising from singlet-triplet transitions[62,63]. Consequently, the atomic magnetoelectric effect in NiPS$_3$ can present potential applications for optical spin read-out devices and two-dimensional opto-spintronic technologies[56,62]. The intriguing ME effect in NiPS$_3$'s quantum-entangled many-body magnetic exciton will also motivate a more detailed examination of other magnetic exciton or spin-orbit exciton systems, such as layered RuCl$_3$[64]. In addition, the electronic and magnetic structure may also be strongly intertwined in similar family systems like FePS$_3$[65], etc.

## 4. Current-induced magnetoelectric effect and spin-orbit torque in Fe$_3$GeTe$_2$

Charge current can generate spin polarisation and so imbalance due to spin-orbit coupling with the mechanisms of spin Hall effect or interfacial Edelstein effect. Such spin polarisation can exert a torque onto the magnetization, called SOT, which conventionally occurs in a ferromagnet/heavy-metal bilayer and depends on the symmetry of the system[66,67]. Fe$_3$GeTe$_2$ represents the earliest discovered vdW ferromagnetic metal with a high Curie temperature of ~200 K and a perpendicular magnetic anisotropy[68,69], and thus has been widely studied for spintronics, particularly the SOT. Conventional protocol of SOT has been applied in a Fe$_3$GeTe$_2$/Pt system[70,71]. At the same time, a different type of gigantic intrinsic SOT[17] in Fe$_3$GeTe$_2$ itself has also been established in recent years. The following paragraphs discuss such gigantic intrinsic SOT by current and the underlying atomic ME effect related to band topology.

Monolayer Fe$_3$GeTe$_2$ has three unique symmetries, including $C_3$, $m_z$ and $m_y$. A theoretical work[18] examined its SOT and found that all the SOT coefficients will be cancelled out regarding symmetry except for a constant $\Gamma_0$. This SOT term can be turned into its free energy formula. Another theory and experiment combined work[17] develops at the same period. The theory found that the SOT term can change the spin-related free energy landscape and reduce the perpendicular magnetic anisotropy by lowering the switching barrier for spin up-to-down transition. In addition, they started from the Kubo formula and derived a formula for calculating the SOT strength. Note that this derivation process handles a question of how much spin accumulation is produced per current, reflecting the microscopic ME effect. The calculated SOT strength, or in the language of the ME effect: the atomic ME coefficient ($\Gamma_0$) is surprisingly found to be as large as ~30 Oe/(mA/μm$^2$). The derived ME formula highlights the effective spin accumulation or the Berry curvature accumulation per current in this system[17]. Therefore, the obtained large ME coefficient[17] is consistent with the previously reported large Berry curvature of FGT's topological bands[72,73]. In the experiment (Fig. 5a-c), a giant coercivity reduction by the current was discovered after subtracting off the Joule heating's contribution using three different but consistent methods. Such substantial coercivity reduction directly results from SOT-reduced magnetic anisotropy and switching energy barrier, ultimately reflecting the strong atomic ME effect. They estimated the atomic ME coefficient from coercivity reduction to be 50 Oe/(mA/μm$^2$), which is close to the theoretical calculation. The ME-related SOT scenario was coherently supported by three approaches: coercivity reduction, theoretical calculations and in-plane angular magnetoresistance. This bulky SOT in Fe$_3$GeTe$_2$ has also been confirmed by a post-coming work from an independent group using the second harmonic technique for electrical transport[74].

Surprisingly, the SOT coefficient is about 100 times larger than that of conventional heavy metals like Pt and Ta[17], demonstrating a strong ME effect that a tiny current can produce a sufficiently large magnetic polarisation. Meanwhile, such SOT occurs in a ferromagnet Fe$_3$GeTe$_2$ without any heavy-metal layer and is deeply rooted in its band topology and Berry curvature[17]. Unlike normal metals or Weyl semimetals[75,76], Fe$_3$GeTe$_2$ has a topological Nodal-line band structure and consequent large Berry curvature[72], and thus eventually large ME effect and SOT magnitude[17]. One interesting question from this work[17] is how to explore a new larger SOT system with band topology. For instance, the larger ME and SOT effect requires large spin polarisation by current due to large spin-orbit coupling or spin Hall effect or the Edelstein effect. These effects require large Berry curvature intrinsically in a material system from the viewpoint of theoretical calculations of Nodal-line materials. Following this logic, one would see huge opportunities to explore the gigantic ME effect in topological Nodal-line materials.

Another insight can be obtained from the symmetry perspective. Since the ME and the SOT coefficients shall conform to the crystalline symmetries, one can explore similar systems containing the same symmetries. Their SOT term will also be simplified in the same way under those symmetries to be incorporated in its free energy expression, enabling the reduction of magnetic anisotropy. Indeed, three reports[77-79] followed the previous pioneering work[17], and used another same-structure material, Fe$_3$GaTe$_2$, to reproduce all the physics and magnetic memory applications of Fe$_3$GeTe$_2$ but at room temperature. We expect more exciting similar systems regarding symmetry in the future, hosting the same ME and SOT physics.

A final issue in this fundamental ME and SOT physics is the inversion symmetry breaking: bulk $Fe_3GeTe_2$ is long thought of as centrosymmetric with layer inversion, but SOT requires inversion symmetry breaking. Previous work[17] addressed this symmetry problem by considering a hidden SOT effect in analogy to the hidden Rashba effect in centrosymmetric systems[80]. Specifically for $Fe_3GeTe_2$, the intralayer magnetic interaction is ~1 eV, about three orders of magnitude higher than the interlayer interaction of ~1 meV and predominates the SOT without considering interlayer interference. Although it nicely explained the symmetry contradiction, it didn't exclude another possibility that inversion symmetry can somehow be broken in actual $Fe_3GeTe_2$ materials. Very recently, the same group discovered the defect-related inversion symmetry breaking by the SHG technique[19], adding another contribution to the intrinsic SOT in bulk $Fe_3GeTe_2$ (Fig. 5d-f). They found that Fe vacancy imbalance can break inversion symmetry mainly along the out-of-plane direction, reducing the space group from centrosymmetric *P*$6_3$/mmc to noncentrosymmetric *P*3m1, consistent with a related structural investigation[81]. The SHG response evolves from a three-fold petal to a more isotropic pattern as the Fe ratio increases from 2.8 to 3, highlighting the defect-breaking layer inversion in bulk $Fe_3GeTe_2$ single crystals.

Meanwhile, the SHG response is independent of temperature and thus purely coming from structural variations. Note that such inversion symmetry breaking is weak, however general it is in similar systems of layer inversion. Moreover, it may affect band topology and other symmetry-related quantum characteristics and hints at other inversion-symmetry-breaking-required phenomena, such as nonlinear optical and electrical response, nonreciprocal transport, and diode effect. $Fe_3GeTe_2$ can also be regarded as the first topological vdW ferromagnetic polar metal ever discovered.

All the above discussions paint a complete physical picture of the atomic ME and intrinsic SOT effect in a single $Fe_3GeTe_2$. These ingredients enable highly efficient magnetic memory with processed developments. In 2021, a novel type of conceptual magnetic memory device[20] has been proposed and systematically demonstrated based on this principle (Fig. 5g-i). The magnetization switching by current is highly efficient and also nonvolatile: the switching current density and power dissipation were reduced by 400 and 4000 times compared to conventional Pt-based devices[20]. Moreover, multi-level states have been readily controlled by tiny current nonvolatilely in the device, which shows around eight states corresponding to 3 bits in a single device, thus enhancing the single-device information capacity and reducing computing cost[20]. Inspired by these two SOT works[17,20], another group succeeded in the real-space imaging of the above switching process by current and the measurement of the domain motion velocity[82]. In 2024, three different works reproduced the same physics and spin memory chip devices using the same-family material $Fe_3GaTe_2$[77-79] at room temperature. Finally, the same group initializing these intrinsic ideas realized the all-vdW three-terminal SOT-MRAM device, where the SOT-writing and TMR-reading process are physically separated to naturally enhance the design flexibility and device endurance, which is closest to an industrial SOT-MRAM architecture. With their fascinating ME and SOT physics and abundant device application demonstrations, the vdW magnets are expected to expand the ME effect towards more exciting spintronic applications.

# 5. Electrical gating on vdW magnets and magnetic control of their electronic properties

In a broad sense, the ME effect can be defined as any electrical control of magnetic properties and, reversely, the magnetic control of electrical properties. Under this context, the most outstanding ME scientific case is the electrical gating of vdW magnets benefiting from its 2D materials' nature. In light of vdW magnetic insulators, $CrI_3$ (Fig. 6a-b) and $CrGeTe_3$ (Fig. 6c-d) are most popular in this aspect. For example, monolayer $CrI_3$ is an Ising-type ferromagnet whose magnetization can be changed by a solid gating voltage[22]. Bilayer $CrI_3$ is an A-type antiferromagnet with an out-of-plane easy axis, where the spins are aligned parallel intralayer for each layer, but anti-parallel interlayer[22,23]. Solid gating a bilayer $CrI_3$ can modulate its magnetization and even realize the antiferromagnetic to ferromagnetic transition under a finite or zero magnetic field[22,23]. The saturated magnetization of $CrGeTe_3$ was also prominently changed in a gating experiment[24]. Moreover, a bipolar tunable magnetization was reported and ascribed to the moment rebalance in the spin-polarised band structures of $CrGeTe_3$[24].

For vdW magnetic metals, ionic gating is much more efficient than solid gating, as reported in many works. For a typical ferromagnetic metal $Fe_3GeTe_2$ (Fig. 6e-f), one group developed the $Al_2O_3$-assisted technique to exfoliate few-layer and monolayer $Fe_3GeTe_2$. It measured their anomalous Hall effect when applying ionic gating down to monolayer[25]. The magnetization and coercive field have been greatly modulated during the electrical gating. Most strikingly, for a four-layer $Fe_3GeTe_2$, the Curie temperature has been enhanced to ~310 K, much larger than even the bulky Curie temperature of ~205 K in bulk $Fe_3GeTe_2$. For another near-room-temperature ferromagnetic metal $Fe_5GeTe_2$ (Fig. 6g-h), ionic gating can even continuously tune its magnetic anisotropy and demonstrate a gradual transition of an out-of-plane easy axis to an in-plane one[26]. It is noted that both carrier doping and electric field can affect magnetism, each of which may be dominant for different experiments. Therefore, careful designs and delicate experiments should be performed regarding this fundamental issue. For instance, the carrier doping effect dominates in some cases, such as in the refs.[22,23,25,26], while the electric field effect plays significant roles in several other cases[24,83]. These four cases are interesting enough to reveal the electrical gating modulation of magnetic properties in vdW magnets, where the saturated magnetization, coercive field, magnetic anisotropy and even the magnetic ground states can be substantially modified by varying the gating voltages. These experiments leverage the charge density variation, electric field effect, Fermi level shift, correlation modification and ultimately magnetic interaction alteration to manipulate the magnetic properties, showcasing the hugely interesting ME effect by electrical means for vdW magnets.

On the other hand, magnetic control of electronic properties has also been reported in several cases with vdW magnets. For example, in bilayer CrSBr, an external magnetic field can induce the exciton energy shift and intensity change due to the spin-flip or spin-canting driven spin configuration change[84]. Theoretically, when spin order changes from antiferromagnetic (AFM) to ferromagnetic (FM) configuration, the interlayer hybridization enhances and leads to the conduction-band-minimum (CBM) and valence-band-maximum (VBM) splitting and consequent band gap reduction. Meanwhile, interlayer hybridization makes relaxation into the lowest-energy bright exciton more efficient and thus makes the PL process more competitive compared to the non-radiative recombination, leading to the PL intensity increase for the FM states. In addition, $CrX_3$ (X=I, Br) is another representative material in which exciton couples with magnetic ordering[85,86]. In this system, PL intensity changes upon applying an external magnetic field, depending on the handedness of the incident and detected light. Such spin-selectable signals can be confirmed not only through emissions but also via current intensity measurements, for instance, using the photovoltaic effect[87]. In particular, $CrI_3$ is a vdW A-type antiferromagnet and thus allows the controllability of its magnetic hysteresis curve depending on whether the layer number is even or odd. Similarly, $MnBi_2Te_4$ features either an axion insulator or a Chern insulator depending on its spin order[88], providing another example of the topological ME effect.

## 6. Outlook to ME effects in vdW quantum magnetic hetero- or homo-structures

Another promising strategy for producing the ME effect is constructing vdW heterostructures by stacking ferromagnetic and ferroelectric materials. In these heterostructures, the broken inversion symmetry at the interface of the two layers can induce a significant ME effect, even if neither material individually exhibits such behaviour. This phenomenon emerges due to the coupling between the ferroelectric polarisation and the magnetic moments at the interface. For example, in the $Fe_3GeTe_2/In_2Se_3$ heterostructure, it has been experimentally demonstrated that applying a voltage can control the magnetic properties of $Fe_3GeTe_2$ through the interface coupling with $In_2Se_3$, thus enabling an electrically tunable ME effect[30]. Two other equally important works include the ferroelectric/ferromagnet heterostructures of $Cr_2Ge_2Te_6$/P(VDF-TrFE)[89] and $Fe_3GaTe_2/CuInP_2S_6$[90], where a voltage applied to the ferroelectric can control the 2D magnetism in cryogenic and room temperature, respectively. Similarly, in other heterostructures, the coupling between ferroelectricity and magnetism has shown potential for nonvolatile memory devices, where electric fields can control magnetic states[28,29]. When designing vdW bilayer structures for the ME effect, lattice mismatch is so critical that a large mismatch can weaken interlayer coupling. One solution to this problem is to create homostructures. A well-known phenomenon related to this is broken inversion symmetry induced by the sliding effect in $MoS_2$ homostructures, which creates a small dipole moment. In this structure, the lattice reconstruction has revealed polarisation in two anti-parallel directions. It is experimentally achieved by stacking layers at specific angles, resulting in a polarisation along the *c*-axis, which can be controlled electrically[91,92].

Similarly, for magnetic vdW materials, stacking engineering can break inversion symmetry even in non-polar structures, thereby inducing a ferroelectric state. In such cases, the interaction between ferroelectric and magnetic ordering can be activated, offering the potential for multiferroic properties. This stacking-engineered ME effect can be applied to study strong intrinsic spin fluctuations associated with two-dimensional magnetism while overcoming material limitations like lattice mismatch. One example is to use this structure as a platform for electrically tuning magnetic domains. This enables the observation of magnetic domains and their boundaries, which is crucial in manifesting topological effects in various materials. Additionally, it aids in analyzing changes in magnetic order due to variations in strain or stacking order, providing insights into ME behaviours at the nanoscale and furthering our understanding of the fundamental physics of two-dimensional magnets.

## 7. Summary

2D vdW magnets provide enormous opportunities for investigating the ME effect at macroscopic and atomic levels, including striking phenomena and underlying mechanisms. In this review, we summarise the ME effect in different vdW magnets, including multiferroic in $NiI_2$, spin-charge correlation in $NiPS_3$, atomic intrinsic SOT in $Fe_3GeTe_2$, gating controlled magnetism in $CrI_3$, $Cr_2Ge_2Te_6$, $Fe_3GeTe_2$, and $Fe_5GeTe_2$, reflecting diverse aspects of the ME effect's manifestation. Finally, we also provide a promising outlook and a few cases on the ME effect in quantum hetero- or homo-structures enabled by assembling these 2D magnets, which can be an exciting new direction along this atomic ME effect topic and corresponding device applications.

**References**

1. Spaldin, N. A., Cheong, S.-W. & Ramesh, R. Multiferroics: Past, present, and future. *Phys. Today* **63**, 38-43 (2010).
2. Kimura, T., Goto, T., Shintani, H., Ishizaka, K., Arima, T.-h. & Tokura, Y. Magnetic control of ferroelectric polarization. *Nature* **426**, 55-58 (2003).
3. Lee, S., Pirogov, A., Kang, M., Jang, K. H., Yonemura, M., Kamiyama, T., Cheong, S. W., Gozzo, F., Shin, N., Kimura, H., Noda, Y. & Park, J. G. Giant magneto-elastic coupling in multiferroic hexagonal manganites. *Nature* **451**, 805-808 (2008).
4. Khomskii, D. Classifying multiferroics: Mechanisms and effects. *Physics* **2**, 20 (2009).
5. Cheong, S.-W. & Mostovoy, M. Multiferroics: a magnetic twist for ferroelectricity. *Nat. Mater.* **6**, 13-20 (2007).
6. Ramesh, R. & Spaldin, N. A. Multiferroics: progress and prospects in thin films. *Nat. Mater.* **6**, 21-29 (2007).
7. Park, J. G. Opportunities and challenges of 2D magnetic van der Waals materials: magnetic graphene? *J. Phys. Condens. Matter* **28**, 301001 (2016).
8. Lee, J. U., Lee, S., Ryoo, J. H., Kang, S., Kim, T. Y., Kim, P., Park, C. H., Park, J. G. & Cheong, H. Ising-Type Magnetic Ordering in Atomically Thin $FePS_3$. *Nano Lett.* **16**, 7433-7438 (2016).
9. Gong, C., Li, L., Li, Z., Ji, H., Stern, A., Xia, Y., Cao, T., Bao, W., Wang, C., Wang, Y., Qiu, Z. Q., Cava, R. J., Louie, S. G., Xia, J. & Zhang, X. Discovery of intrinsic ferromagnetism in two-dimensional van der Waals crystals. *Nature* **546**, 265-269 (2017).
10. Huang, B., Clark, G., Navarro-Moratalla, E., Klein, D. R., Cheng, R., Seyler, K. L., Zhong, D., Schmidgall, E., McGuire, M. A., Cobden, D. H., Yao, W., Xiao, D., Jarillo-Herrero, P. & Xu, X. Layer-dependent ferromagnetism in a van der Waals crystal down to the monolayer limit. *Nature* **546**, 270-273 (2017).
11. Burch, K. S., Mandrus, D. & Park, J. G. Magnetism in two-dimensional van der Waals materials. *Nature* **563**, 47-52 (2018).
12. Ju, H., Lee, Y., Kim, K. T., Choi, I. H., Roh, C. J., Son, S., Park, P., Kim, J. H., Jung, T. S., Kim, J. H., Kim, K. H., Park, J. G. & Lee, J. S. Possible Persistence of Multiferroic Order down to Bilayer Limit of van der Waals Material $NiI_2$. *Nano Lett.* **21**, 5126-5132 (2021).
13. Song, Q., Occhialini, C. A., Ergecen, E., Ilyas, B., Amoroso, D., Barone, P., Kapeghian, J., Watanabe, K., Taniguchi, T., Botana, A. S., Picozzi, S., Gedik, N. & Comin, R. Evidence for a single-layer van der Waals multiferroic. *Nature* **602**, 601-605 (2022).
14. Son, S., Lee, Y., Kim, J. H., Kim, B. H., Kim, C., Na, W., Ju, H., Park, S., Nag, A., Zhou, K. J., Son, Y. W., Kim, H., Noh, W. S., Park, J. H., Lee, J. S., Cheong, H., Kim, J. H. & Park, J. G. Multiferroic-Enabled Magnetic-Excitons in 2D Quantum-Entangled Van der Waals Antiferromagnet $NiI_2$. *Adv. Mater.* **34**, 2109144 (2022).
15. Kang, S., Kim, K., Kim, B. H., Kim, J., Sim, K. I., Lee, J.-U., Lee, S., Park, K., Yun, S., Kim, T., Nag, A., Walters, A., Garcia-Fernandez, M., Li, J., Chapon, L., Zhou, K.-J., Son, Y.-W., Kim, J. H., Cheong, H. & Park, J.-G. Coherent many-body exciton in van der Waals antiferromagnet $NiPS_3$. *Nature* **583**, 785-789 (2020).
16. Kim, S. Y., Kim, T. Y., Sandilands, L. J., Sinn, S., Lee, M. C., Son, J., Lee, S., Choi, K. Y., Kim, W., Park, B. G., Jeon, C., Kim, H. D., Park, C. H., Park, J. G., Moon, S. J. & Noh, T. W. Charge-Spin Correlation in van der Waals Antiferromagnet $NiPS_3$. *Phys.*

*Rev. Lett.* **120**, 136402 (2018).
17. Zhang, K., Han, S., Lee, Y., Coak, M. J., Kim, J., Hwang, I., Son, S., Shin, J., Lim, M., Jo, D., Kim, K., Kim, D., Lee, H.-W. & Park, J.-G. Gigantic current control of coercive field and magnetic memory based on nm-thin ferromagnetic van der Waals $Fe_3GeTe_2$. *Adv. Mater.* **33**, 2004110 (2021).
18. Johansen, Ø., Risinggård, V., Sudbø, A., Linder, J. & Brataas, A. Current Control of Magnetism in Two-Dimensional $Fe_3GeTe_2$. *Phys. Rev. Lett.* **122**, 217203 (2019).
19. Zhang, K.-X., Ju, H., Kim, H., Cui, J., Keum, J., Park, J.-G. & Lee, J. S. Broken inversion symmetry in van der Waals topological ferromagnetic metal iron germanium telluride. *Adv. Mater.* **36**, 2312824 (2024).
20. Zhang, K., Lee, Y., Coak, M. J., Kim, J., Son, S., Hwang, I., Ko, D. S., Oh, Y., Jeon, I., Kim, D., Zeng, C., Lee, H.-W. & Park, J.-G. Highly efficient nonvolatile magnetization switching and multi-level states by current in single van der Waals topological ferromagnet $Fe_3GeTe_2$. *Adv. Funct. Mater.* **31**, 2105992 (2021).
21. Cui, J., Zhang, K.-X. & Park, J.-G. All van der Waals Three-Terminal SOT-MRAM Realized by Topological Ferromagnet $Fe_3GeTe_2$. *Adv. Electron. Mater.* **10**, 2400041 (2024).
22. Jiang, S., Li, L., Wang, Z., Mak, K. F. & Shan, J. Controlling magnetism in 2D $CrI_3$ by electrostatic doping. *Nat. Nanotechnol.* **13**, 549-553 (2018).
23. Huang, B., Clark, G., Klein, D. R., MacNeill, D., Navarro-Moratalla, E., Seyler, K. L., Wilson, N., McGuire, M. A., Cobden, D. H., Xiao, D., Yao, W., Jarillo-Herrero, P. & Xu, X. Electrical control of 2D magnetism in bilayer $CrI_3$. *Nat. Nanotechnol.* **13**, 544-548 (2018).
24. Wang, Z., Zhang, T., Ding, M., Dong, B., Li, Y., Chen, M., Li, X., Huang, J., Wang, H., Zhao, X., Li, Y., Li, D., Jia, C., Sun, L., Guo, H., Ye, Y., Sun, D., Chen, Y., Yang, T., Zhang, J., Ono, S., Han, Z. & Zhang, Z. Electric-field control of magnetism in a few-layered van der Waals ferromagnetic semiconductor. *Nat. Nanotechnol.* **13**, 554-559 (2018).
25. Deng, Y., Yu, Y., Song, Y., Zhang, J., Wang, N. Z., Sun, Z., Yi, Y., Wu, Y. Z., Wu, S., Zhu, J., Wang, J., Chen, X. H. & Zhang, Y. Gate-tunable room-temperature ferromagnetism in two-dimensional $Fe_3GeTe_2$. *Nature* **563**, 94-99 (2018).
26. Tang, M., Huang, J., Qin, F., Zhai, K., Ideue, T., Li, Z., Meng, F., Nie, A., Wu, L., Bi, X., Zhang, C., Zhou, L., Chen, P., Qiu, C., Tang, P., Zhang, H., Wan, X., Wang, L., Liu, Z., Tian, Y., Iwasa, Y. & Yuan, H. Continuous manipulation of magnetic anisotropy in a van der Waals ferromagnet via electrical gating. *Nat. Electron.* **6**, 28-36 (2023).
27. Zheng, G., Xie, W.-Q., Albarakati, S., Algarni, M., Tan, C., Wang, Y., Peng, J., Partridge, J., Farrar, L., Yi, J., Xiong, Y., Tian, M., Zhao, Y.-J. & Wang, L. Gate-Tuned Interlayer Coupling in van der Waals Ferromagnet $Fe_3GeTe_2$ Nanoflakes. *Phys. Rev. Lett.* **125**, 047202 (2020).
28. Huang, X., Li, G., Chen, C., Nie, X., Jiang, X. & Liu, J.-M. Interfacial coupling induced critical thickness for the ferroelectric bistability of two-dimensional ferromagnet/ferroelectric van der Waals heterostructures. *Phys. Rev. B* **100**, 235445 (2019).
29. Cao, L., Deng, X., Zhou, G., Liang, S.-J., Nguyen, C. V., Ang, L. & Ang, Y. S. Multiferroic van der Waals heterostructure $FeCl_2$/$Sc_2CO_2$: Nonvolatile electrically switchable electronic and spintronic properties. *Phys. Rev. B* **105**, 165302 (2022).
30. Eom, J., Lee, I. H., Kee, J. Y., Cho, M., Seo, J., Suh, H., Choi, H.-J., Sim, Y., Chen, S.,

Chang, H. J., Baek, S.-H., Petrovic, C., Ryu, H., Jang, C., Kim, Y. D., Yang, C.-H., Seong, M.-J., Lee, J. H., Park, S. Y. & Choi, J. W. Voltage control of magnetism in $Fe_{3-x}GeTe_2/In_2Se_3$ van der Waals ferromagnetic/ferroelectric heterostructures. *Nat. Commun.* **14**, 5605 (2023).

31. Kurumaji, T., Seki, S., Ishiwata, S., Murakawa, H., Kaneko, Y. & Tokura, Y. Magnetoelectric responses induced by domain rearrangement and spin structural change in triangular-lattice helimagnets $NiI_2$ and $CoI_2$. *Phys. Rev. B* **87**, 014429 (2013).
32. Kurumaji, T., Seki, S., Ishiwata, S., Murakawa, H., Tokunaga, Y., Kaneko, Y. & Tokura, Y. Magnetic-Field Induced Competition of Two Multiferroic Orders in a Triangular-Lattice Helimagnet $MnI_2$. *Phys. Rev. Lett.* **106**, 167206 (2011).
33. Tokunaga, Y., Okuyama, D., Kurumaji, T., Arima, T., Nakao, H., Murakami, Y., Taguchi, Y. & Tokura, Y. Multiferroicity in $NiBr_2$ with long-wavelength cycloidal spin structure on a triangular lattice. *Phys. Rev. B* **84**, 060406 (2011).
34. Amini, M., Fumega, A. O., González-Herrero, H., Vaňo, V., Kezilebieke, S., Lado, J. L. & Liljeroth, P. Atomic-Scale Visualization of Multiferroicity in Monolayer $NiI_2$. *Adv. Mater.* **36**, 2311342 (2024).
35. Liu, N., Wang, C., Yan, C., Xu, C., Hu, J., Zhang, Y. & Ji, W. Competing multiferroic phases in monolayer and few-layer $NiI_2$. *Phys. Rev. B* **109**, 195422 (2024).
36. van der Laan, G., Zaanen, J., Sawatzky, G. A., Karnatak, R. & Esteva, J. Comparison of x-ray absorption with x-ray photoemission of nickel dihalides and NiO. *Phys. Rev. B* **33**, 4253-4263 (1986).
37. Kim, J. H., Jung, T. S., Lee, Y., Kim, C., Park, J.-G. & Kim, J. H. Terahertz evidence of electromagnon excitations in the multiferroic van der Waals insulator $NiI_2$. *Phys. Rev. B* **108**, 064414 (2023).
38. Gao, F. Y., Peng, X., Cheng, X., Viñas Boström, E., Kim, D. S., Jain, R. K., Vishnu, D., Raju, K., Sankar, R., Lee, S.-F., Sentef, M. A., Kurumaji, T., Li, X., Tang, P., Rubio, A. & Baldini, E. Giant chiral magnetoelectric oscillations in a van der Waals multiferroic. *Nature* **632**, 273-279 (2024).
39. Masuda, R., Kaneko, Y., Tokura, Y. & Takahashi, Y. Electric field control of natural optical activity in a multiferroic helimagnet. *Science* **372**, 496-500 (2021).
40. Liu, C., Ren, W. & Picozzi, S. Spin-chirality-driven multiferroicity in van der waals monolayers. *Phys. Rev. Lett.* **132**, 086802 (2024).
41. Sødequist, J. & Olsen, T. Type II multiferroic order in two-dimensional transition metal halides from first principles spin-spiral calculations. *2D Mater.* **10**, 035016 (2023).
42. Wang, X., Shang, Z., Zhang, C., Kang, J., Liu, T., Wang, X., Chen, S., Liu, H., Tang, W., Zeng, Y. J., Guo, J., Cheng, Z., Liu, L., Pan, D., Tong, S., Wu, B., Xie, Y., Wang, G., Deng, J., Zhai, T., Deng, H. X., Hong, J. & Zhao, J. Electrical and magnetic anisotropies in van der Waals multiferroic $CuCrP_2S_6$. *Nat. Commun.* **14**, 840 (2023).
43. Aoki, S., Dong, Y., Wang, Z., Huang, X. S., Itahashi, Y. M., Ogawa, N., Ideue, T. & Iwasa, Y. Giant Modulation of the Second Harmonic Generation by Magnetoelectricity in Two-Dimensional Multiferroic $CuCrP_2S_6$. *Adv. Mater.* **36**, 2312781 (2024).
44. Wildes, A. R., Simonet, V., Ressouche, E., Mcintyre, G. J., Avdeev, M., Suard, E., Kimber, S. A., Lançon, D., Pepe, G., Moubaraki, B. & Hicks, T. J. Magnetic structure of the quasi-two-dimensional antiferromagnet $NiPS_3$. *Phys. Rev. B* **92**, 224408 (2015).
45. Kim, K., Lim, S. Y., Lee, J. U., Lee, S., Kim, T. Y., Park, K., Jeon, G. S., Park, C. H., Park, J. G. & Cheong, H. Suppression of magnetic ordering in XXZ-type antiferromagnetic monolayer NiPS$_3$. *Nat. Commun.* **10**, 345 (2019).

46. Yan, M., Jin, Y., Wu, Z., Tsaturyan, A., Makarova, A., Smirnov, D., Voloshina, E. & Dedkov, Y. Correlations in the Electronic Structure of van der Waals NiPS$_3$ Crystals: An X-ray Absorption and Resonant Photoelectron Spectroscopy Study. *J. Phys. Chem. Lett.* **12**, 2400-2405 (2021).
47. He, W., Shen, Y., Wohlfeld, K., Sears, J., Li, J., Pelliciari, J., Walicki, M., Johnston, S., Baldini, E., Bisogni, V., Mitrano, M. & Dean, M. P. M. Magnetically propagating Hund's exciton in van der Waals antiferromagnet NiPS$_3$. *Nat. Commun.* **15**, 3496 (2024).
48. Khomskii, D. Unusual valence, negative charge-transfer gaps and self-doping in transition-metal compounds. *Lith. J. Phys.* **37**, 65 (1997).
49. Belvin, C. A., Baldini, E., Ozel, I. O., Mao, D., Po, H. C., Allington, C. J., Son, S., Kim, B. H., Kim, J., Hwang, I., Kim, J. H., Park, J. G., Senthil, T. & Gedik, N. Exciton-driven antiferromagnetic metal in a correlated van der Waals insulator. *Nat. Commun.* **12**, 4837 (2021).
50. Kim, H. J. & Kim, K.-S. Microscopic origin of local electric polarization in NiPS$_3$. *New J. Phys.* **25**, 083029 (2023).
51. Zhang, F. & Rice, T. Effective Hamiltonian for the superconducting Cu oxides. *Phys. Rev. B* **37**, 3759 (1988).
52. Maekawa, S., Tohyama, T., Barnes, S. E., Ishihara, S., Koshibae, W. & Khaliullin, G. Physics of Transition Metal Oxides (Springer-Verlag, Heidelberg, 2004).
53. Klaproth, T., Aswartham, S., Shemerliuk, Y., Selter, S., Janson, O., van den Brink, J., Buchner, B., Knupfer, M., Pazek, S., Mikhailova, D., Efimenko, A., Hayn, R., Savoyant, A., Gubanov, V. & Koitzsch, A. Origin of the Magnetic Exciton in the van der Waals Antiferromagnet NiPS$_3$. *Phys. Rev. Lett.* **131**, 256504 (2023).
54. Ergeçen, E., Ilyas, B., Mao, D., Po, H. C., Yilmaz, M. B., Kim, J., Park, J.-G., Senthil, T. & Gedik, N. Magnetically brightened dark electron-phonon bound states in a van der Waals antiferromagnet. *Nat. Commun.* **13**, 98 (2022).
55. Wang, X., Cao, J., Lu, Z., Cohen, A., Kitadai, H., Li, T., Tan, Q., Wilson, M., Lui, C. H., Smirnov, D., Sharifzadeh, S. & Ling, X. Spin-induced linear polarization of photoluminescence in antiferromagnetic van der Waals crystals. *Nat. Mater.* **20**, 964 (2021).
56. Hwangbo, K., Zhang, Q., Jiang, Q., Wang, Y., Fonseca, J., Wang, C., Diederich, G. M., Gamelin, D. R., Xiao, D., Chu, J.-H., Yao, W. & Xu, X. Highly anisotropic excitons and multiple phonon bound states in a van der Waals antiferromagnetic insulator. *Nat. Nanotechnol.* **16**, 655 (2021).
57. Tan, Q., Luo, W., Li, T., Cao, J., Kitadai, H., Wang, X. & Ling, X. Charge-transfer-enhanced d–d emission in antiferromagnetic NiPS$_3$. *Applied Physics Reviews* **9**, 041406 (2022).
58. Dirnberger, F., Bushati, R., Datta, B., Kumar, A., MacDonald, A. H., Baldini, E. & Menon, V. M. Spin-correlated exciton-polaritons in a van der Waals magnet. *Nat. Nanotechnol.* **17**, 1060-1064 (2022).
59. Kim, D. S., Huang, D., Guo, C., Li, K., Rocca, D., Gao, F. Y., Choe, J., Lujan, D., Wu, T. H., Lin, K. H., Baldini, E., Yang, L., Sharma, S., Kalaivanan, R., Sankar, R., Lee, S. F., Ping, Y. & Li, X. Anisotropic Excitons Reveal Local Spin Chain Directions in a van der Waals Antiferromagnet. *Adv. Mater.* **35**, 2206585 (2023).
60. Kim, J. K., Na, W., Kim, J., Park, P., Zhang, K., Hwang, I., Son, Y.-W., Kim, J. H., Cheong, H. & Park, J.-G. Rapid Suppression of Quantum Many-Body Magnetic

Exciton in Doped van der Waals Antiferromagnet (Ni,Cd)PS$_3$. *Nano Lett.* **23**, 10189-10195 (2023).

61. Jana, D., Kapuscinski, P., Mohelsky, I., Vaclavkova, D., Breslavetz, I., Orlita, M., Faugeras, C. & Potemski, M. Magnon gap excitations and spin-entangled optical transition in the van der Waals antiferromagnet NiPS$_3$. *Phys. Rev. B* **108**, 115149 (2023).
62. Wang, X., Tan, Q., Li, T., Lu, Z., Cao, J., Ge, Y., Zhao, L., Tang, J., Kitadai, H., Guo, M., Li, Y. M., Xu, W., Cheng, R., Smirnov, D. & Ling, X. Unveiling the spin evolution in van der Waals antiferromagnets via magneto-exciton effects. *Nat. Commun.* **15**, 8011 (2024).
63. Song, F., Lv, Y., Sun, Y. J., Pang, S., Chang, H., Guan, S., Lai, J. M., Wang, X. J., Wu, B., Hu, C., Yuan, Z. & Zhang, J. Manipulation of anisotropic Zhang-Rice exciton in NiPS$_3$ by magnetic field. *Nat. Commun.* **15**, 7841 (2024).
64. Lee, J.-H., Choi, Y., Do, S.-H., Kim, B. H., Seong, M.-J. & Choi, K.-Y. Multiple spin-orbit excitons in α-RuCl$_3$ from bulk to atomically thin layers. *npj Quantum Mater.* **6**, 43 (2021).
65. Koitzsch, A., Klaproth, T., Selter, S., Shemerliuk, Y., Aswartham, S., Janson, O., Bnchner, B. & Knupfer, M. Intertwined electronic and magnetic structure of the van-der-Waals antiferromagnet Fe$_2$P$_2$S$_6$. *npj Quantum Mater.* **8**, 27 (2023).
66. Manchon, A., Železný, J., Miron, I. M., Jungwirth, T., Sinova, J., Thiaville, A., Garello, K. & Gambardella, P. Current-induced spin-orbit torques in ferromagnetic and antiferromagnetic systems. *Rev. Mod. Phys.* **91**, 035004 (2019).
67. Kurebayashi, H., Garcia, J. H., Khan, S., Sinova, J. & Roche, S. Magnetism, symmetry and spin transport in van der Waals layered systems. *Nat. Rev. Phys.* **4**, 150-166 (2022).
68. Fei, Z., Huang, B., Malinowski, P., Wang, W., Song, T., Sanchez, J., Yao, W., Xiao, D., Zhu, X., May, A. F., Wu, W., Cobden, D. H., Chu, J. H. & Xu, X. Two-dimensional itinerant ferromagnetism in atomically thin Fe$_3$GeTe$_2$. *Nat. Mater.* **17**, 778-782 (2018).
69. Tan, C., Lee, J., Jung, S. G., Park, T., Albarakati, S., Partridge, J., Field, M. R., McCulloch, D. G., Wang, L. & Lee, C. Hard magnetic properties in nanoflake van der Waals Fe$_3$GeTe$_2$. *Nat. Commun.* **9**, 1554 (2018).
70. Wang, X., Tang, J., Xia, X., He, C., Zhang, J., Liu, Y., Wan, C., Fang, C., Guo, C., Yang, W., Guang, Y., Zhang, X., Xu, H., Wei, J., Liao, M., Lu, X., Feng, J., Li, X., Peng, Y., Wei, H., Yang, R., Shi, D., Zhang, X., Han, Z., Zhang, Z., Zhang, G., Yu, G. & Han, X. Current-driven magnetization switching in a van der Waals ferromagnet Fe$_3$GeTe$_2$. *Sci. Adv.* **5**, eaaw8904 (2019).
71. Alghamdi, M., Lohmann, M., Li, J., Jothi, P. R., Shao, Q., Aldosary, M., Su, T., Fokwa, B. & Shi, J. Highly Efficient Spin-Orbit Torque and Switching of Layered Ferromagnet Fe$_3$GeTe$_2$. *Nano Lett.* **19**, 4400-4405 (2019).
72. Kim, K., Seo, J., Lee, E., Ko, K. T., Kim, B. S., Jang, B. G., Ok, J. M., Lee, J., Jo, Y. J., Kang, W., Shim, J. H., Kim, C., Yeom, H. W., Il Min, B., Yang, B. J. & Kim, J. S. Large anomalous Hall current induced by topological nodal lines in a ferromagnetic van der Waals semimetal. *Nat. Mater.* **17**, 794-799 (2018).
73. Xu, J., Phelan, W. A. & Chien, C. L. Large Anomalous Nernst Effect in a van der Waals Ferromagnet Fe$_3$GeTe$_2$. *Nano Lett.* **19**, 8250-8254 (2019).
74. Martin, F., Lee, K., Schmitt, M., Liedtke, A., Shahee, A., Simensen, H. T., Scholz, T., Saunderson, T. G., Go, D., Gradhand, M., Mokrousov, Y., Denneulin, T., Kovács, A., Lotsch, B., Brataas, A. & Kläui, M. Strong bulk spin-orbit torques quantified in the van der Waals ferromagnet Fe$_3$GeTe$_2$. *Mater. Res. Lett.* **11**, 84-89 (2022).

75. Zhang, K., Du, Y., Wang, P., Wei, L., Li, L., Zhang, Q., Qin, W., Lin, Z., Cheng, B., Wang, Y., Xu, H., Fan, X., Zhe Sun, Wan, X. & Zeng, C. Butterfly-Like Anisotropic Magnetoresistance and Angle-Dependent Berry Phase in a Type-II Weyl Semimetal WP$_2$. *Chin. Phys. Lett.* **37**, 090301 (2020).
76. Zhang, K., Du, Y., Qi, Z., Cheng, B., Fan, X., Wei, L., Li, L., Wang, D., Yu, G., Hu, S., Sun, C., Huang, Z., Chu, J., Wan, X. & Zeng, C. Room-Temperature Anisotropic Plasma Mirror and Polarization-Controlled Optical Switch Based on Type-II Weyl Semimetal WP$_2$. *Phys. Rev. Appl.* **13**, 014058 (2020).
77. Zhang, G., Wu, H., Yang, L., Jin, W., Xiao, B., Zhang, W. & Chang, H. Room-temperature Highly-Tunable Coercivity and Highly-Efficient Multi-States Magnetization Switching by Small Current in Single 2D Ferromagnet Fe$_3$GaTe$_2$. *ACS Mater. Lett.* **6**, 482-488 (2024).
78. Yan, S., Tian, S., Fu, Y., Meng, F., Li, Z., Lei, H., Wang, S. & Zhang, X. Highly Efficient Room-Temperature Nonvolatile Magnetic Switching by Current in Fe$_3$GaTe$_2$ Thin Flakes. *Small* **20**, 2311430 (2024).
79. Deng, Y., Wang, M., Xiang, Z., Zhu, K., Hu, T., Lu, L., Wang, Y., Ma, Y., Lei, B. & Chen, X. Room-Temperature Highly Efficient Nonvolatile Magnetization Switching by Current in van der Waals Fe$_3$GaTe$_2$ Devices. *Nano Lett.* **24**, 9302-9310 (2024).
80. Zhang, X., Liu, Q., Luo, J.-W., Freeman, A. J. & Zunger, A. Hidden spin polarization in inversion-symmetric bulk crystals. *Nat. Phys.* **10**, 387-393 (2014).
81. Chakraborty, A., Srivastava, A. K., Sharma, A. K., Gopi, A. K., Mohseni, K., Ernst, A., Deniz, H., Hazra, B. K., Das, S., Sessi, P., Kostanovskiy, I., Ma, T., Meyerheim, H. L. & Parkin, S. S. P. Magnetic Skyrmions in a Thickness Tunable 2D Ferromagnet from a Defect Driven Dzyaloshinskii-Moriya Interaction. *Adv. Mater.* **34**, e2108637 (2022).
82. Robertson, I. O., Tan, C., Scholten, S. C., Healey, A. J., Abrahams, G. J., Zheng, G., Manchon, A., Wang, L. & Tetienne, J.-P. Imaging current control of magnetization in Fe$_3$GeTe$_2$ with a widefield nitrogen-vacancy microscope. *2D Mater.* **10**, 015023 (2022).
83. Jiang, S., Shan, J. & Mak, K. F. Electric-field switching of two-dimensional van der Waals magnets. *Nat. Mater.* **17**, 406-410 (2018).
84. Wilson, N. P., Lee, K., Cenker, J., Xie, K., Dismukes, A. H., Telford, E. J., Fonseca, J., Sivakumar, S., Dean, C., Cao, T., Roy, X., Xu, X. & Zhu, X. Interlayer electronic coupling on demand in a 2D magnetic semiconductor. *Nat. Mater.* **20**, 1657-1662 (2021).
85. Jin, W., Kim, H. H., Ye, Z., Ye, G., Rojas, L., Luo, X., Yang, B., Yin, F., Horng, J. S. A., Tian, S., Fu, Y., Xu, G., Deng, H., Lei, H., Tsen, A. W., Sun, K., He, R. & Zhao, L. Observation of the polaronic character of excitons in a two-dimensional semiconducting magnet CrI$_3$. *Nat. Commun.* **11**, 4780 (2020).
86. Grzeszczyk, M., Acharya, S., Pashov, D., Chen, Z., Vaklinova, K., van Schilfgaarde, M., Watanabe, K., Taniguchi, T., Novoselov, K. & Katsnelson, M. Strongly Correlated Exciton‐Magnetization System for Optical Spin Pumping in CrBr$_3$ and CrI$_3$. *Adv. Mater.* **35**, 2209513 (2023).
87. Song, T., Anderson, E., Tu, M. W., Seyler, K., Taniguchi, T., Watanabe, K., McGuire, M. A., Li, X., Cao, T., Xiao, D., Yao, W. & Xu, X. Spin photovoltaic effect in magnetic van der Waals heterostructures. *Sci. Adv.* **7**, eabg8094 (2021).
88. Liu, C., Wang, Y., Li, H., Wu, Y., Li, Y., Li, J., He, K., Xu, Y., Zhang, J. & Wang, Y. Robust axion insulator and Chern insulator phases in a two-dimensional antiferromagnetic topological insulator. *Nat. Mater.* **19**, 522-527 (2020).
89. Liang, S. C., Xie, T., Blumenschein, N. A., Zhou, T., Ersevim, T., Song, Z. H., Liang, J. R., Susner, M. A., Conner, B. S., Gong, S. J., Wang, J. P., Ouyang, M., Zutic, I., Friedman, A. L., Zhang, X. & Gong, C. Small-voltage multiferroic control of two-dimensional magnetic insulators. *Nat. Electron.* **6**, 199-205 (2023).
90. Wu, Y., k Sofer, Z., Karuppasamy, M. & Wang, W. Room-temperature Ferroelectric Control of 2D Layered Magnetism. *arXiv 2406.13859 (2024)*.
91. Wang, X., Yasuda, K., Zhang, Y., Liu, S., Watanabe, K., Taniguchi, T., Hone, J., Fu, L. & Jarillo-Herrero, P. Interfacial ferroelectricity in rhombohedral-stacked bilayer transition metal dichalcogenides. *Nat. Nanotechnol.* **17**, 367-371 (2022).
92. Vizner Stern, M., Waschitz, Y., Cao, W., Nevo, I., Watanabe, K., Taniguchi, T., Sela, E., Urbakh, M., Hod, O. & Ben Shalom, M. Interfacial ferroelectricity by van der Waals sliding. *Science* **372**, 1462-1466 (2021).

**Acknowledgements**

We acknowledge the numerous discussions we had with our collaborators, which led us to improve our understanding of this problem. We are particularly indebted to Suhan Son and Jongseok Lee for their generous help. This work was supported by the Samsung Science & Technology Foundation (Grant No. SSTF-BA2101-05). One of the authors (J.-G.P.) is partly funded by the Leading Researcher Program of the National Research Foundation of Korea (Grant No. 2020R1A3B2079375). B.H.K. was supported by the Institute for Basic Science in the Republic of Korea (Project No. IBS-R024-D1).

**Author contributions**

K.-X.Z. and J.-G.P. initiated and supervised the review paper. J.-G.P. formulated the overall structure of the article. G.P. worked on the outlook part of heterostructures and the section on magnetic control of electronic properties. Y.L. worked on the $NiI_2$ part. B.H.K. worked on the $NiPS_3$ part. K.-X.Z. worked on the $Fe_3GeTe_2$ section and other remaining parts. K.-X.Z. and J.-G.P. wrote and revised the manuscript with contributions from all the co-authors.

**Competing interests**

The authors declare no competing interests.

**Figures**

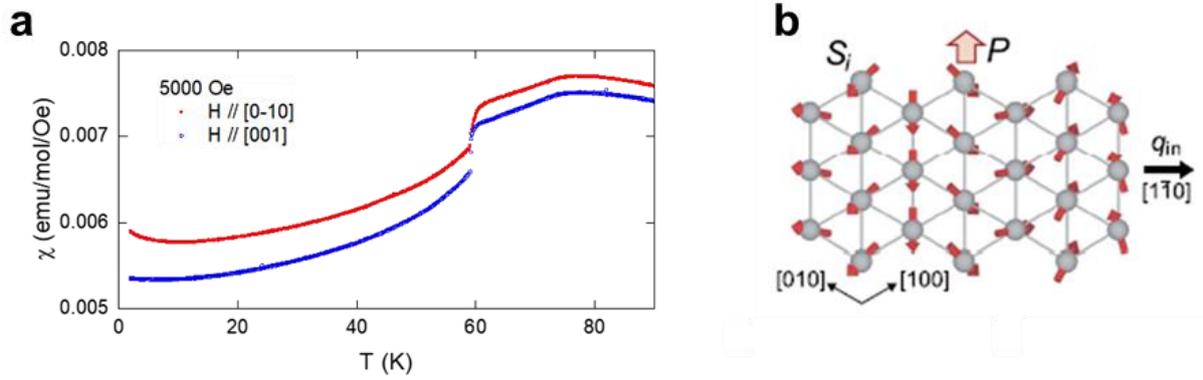

**Fig.1. Multiferroic properties of NiI$_2$. a**, Magnetic susceptibility of NiI$_2$ with two successive phase transitions: the paramagnetic-to-antiferromagnetic transition at $T_{N1}$ = 76 K and the helimagnetic transition at $T_{N2}$ = 59.5 K. **b**, In-plane ferroelectricity (P) with helimagnetic spin order. (Fig. 1a is reproduced from Fig. 1e in Ref[14]; Fig. 1b is reproduced from Fig. 1b in Ref[12].)

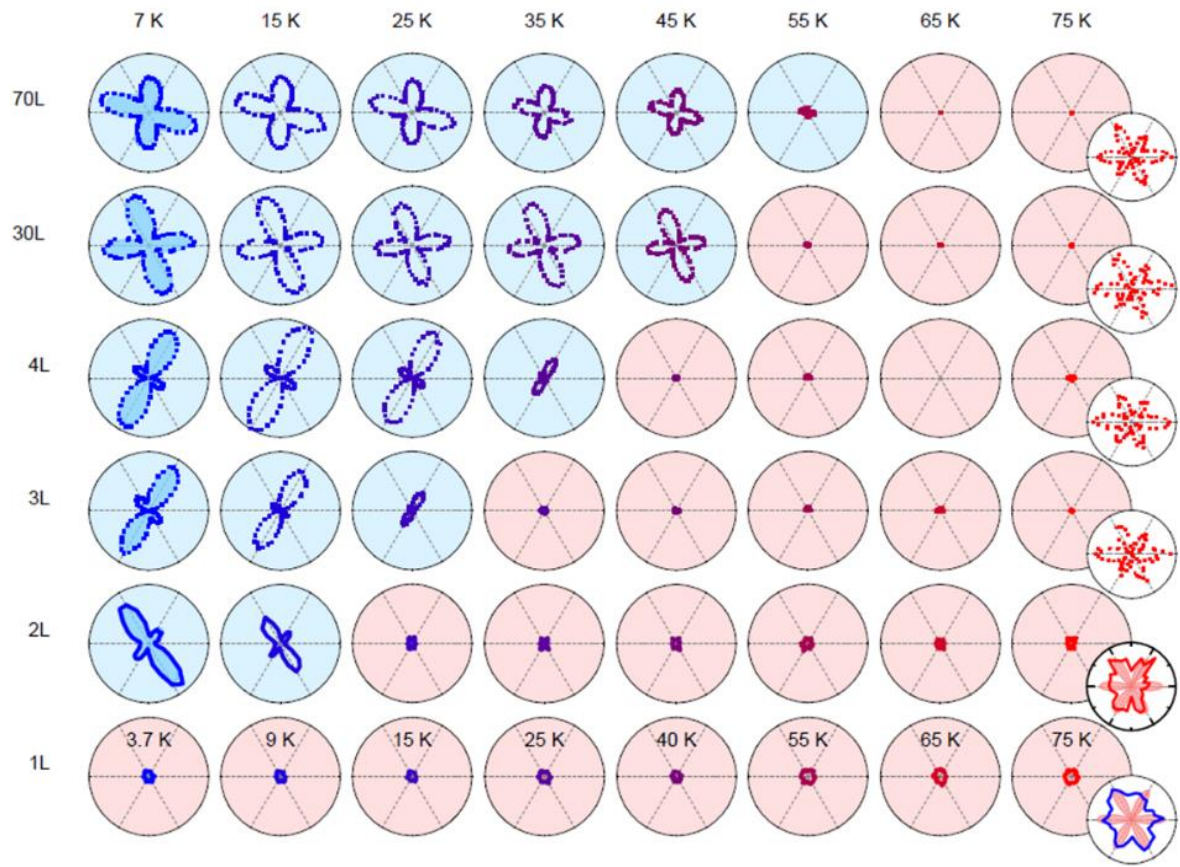

**Fig. 2. Thickness-dependent multiferroic order in NiI$_2$**. It shows different rotational anisotropy of the second-harmonic generation (SHG) pattern as a function of NiI$_2$ thickness from 70 layers (70L) to monolayer (1L). Red and blue background colours highlight the transition to the polar state upon decreasing temperature. The red colour represents the non-ferroelectric state with a minute SHG response of six petals. In comparison, the blue colour represents the ferroelectric state with a significant SHG response of four petals. The ferroelectric state is maintained by an exfoliated NiI$_2$ bilayer on the SiO$_2$/Si substrate. (Fig. 2 is remade from Fig. 3 in Ref[12].)

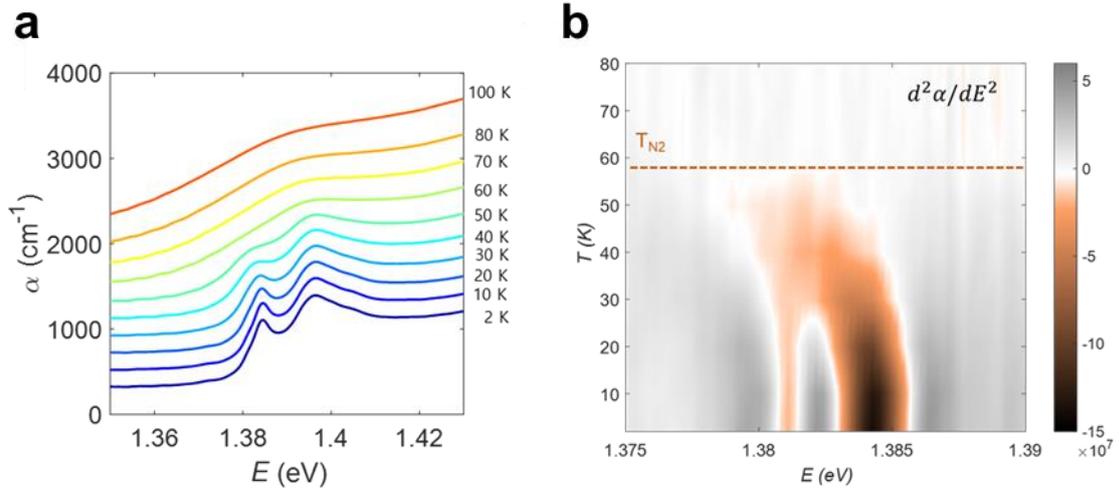

**Fig. 3. Optical absorption properties of NiI$_2$. a**, Optical absorption of NiI$_2$ with magnetic exciton at $E_{hv}$ = 1.384 eV and two-magnon side-band peak at $E_{hv}$ = 1.396 eV below $T_{N2}$. **b**, Derivative optical absorption shows a clear correlation between multiferroicity and the magnetic exciton. (Fig. 3a,b is reproduced from Fig. 2a,c in Ref[14].)

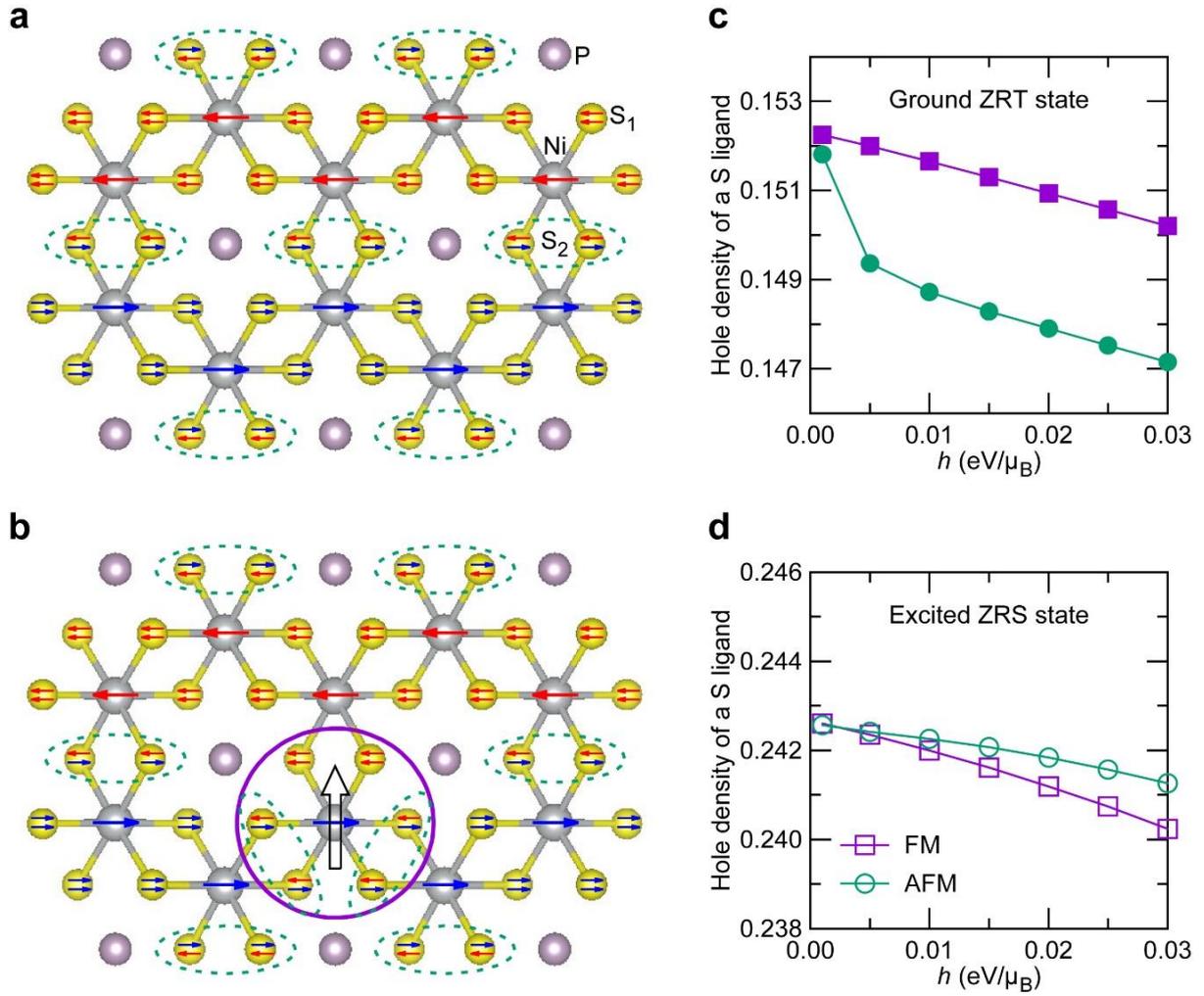

**Fig. 4. Schematic diagrams of spin configurations and hole density of a S ligand in a $Ni_2S_{10}$ cluster. a-b**, Schematic diagrams of spin configurations in (**a**) the Zhang-Rice triplet (ZRT) lattice and (**b**) the Zhang-Rice singlet (ZRS) excited lattice under zigzag magnetic order in $NiPS_3$. Large (small) red and blue arrows represent the spin directions of Ni $3d$ (S $3p$) holes, respectively. Green dotted ellipsoids highlight $S_2$ sites, where two ligand holes exhibit anti-parallel spins, while $S_1$ sites, where the two holes have parallel spins, are depicted for the other S atoms. The magenta circle indicates the locations of ZRS excitations. Due to spin flipping in one of the hole orbitals at the surrounding S sites, the transition from ZRT to ZRS states leads to polarisation reversal. It induces electric dipoles, with the direction indicated by the open arrow inside the magenta circle. **c-d**, Hole density of an S ligand as a function of the long-range ordering field strength $h$, for the ground ZRT state (**c**) and the excited ZRS state (**d**), respectively, as calculated in a $Ni_2S_{10}$ cluster (see details in Ref[15]). FM (AFM) refers to cases where the applied long-range ordering fields at the two Ni sites are parallel (anti-parallel). The differences in hole density between the FM and AFM cases in (**c**) and (**d**) indicate the charge modulations of self-doped ligand holes at $S_1$ and $S_2$ sites in the ground ZRT state and the polarisation reversal in the excited ZRS state, respectively. (Fig. 4 is remade from Fig. S9 in Ref[15].)

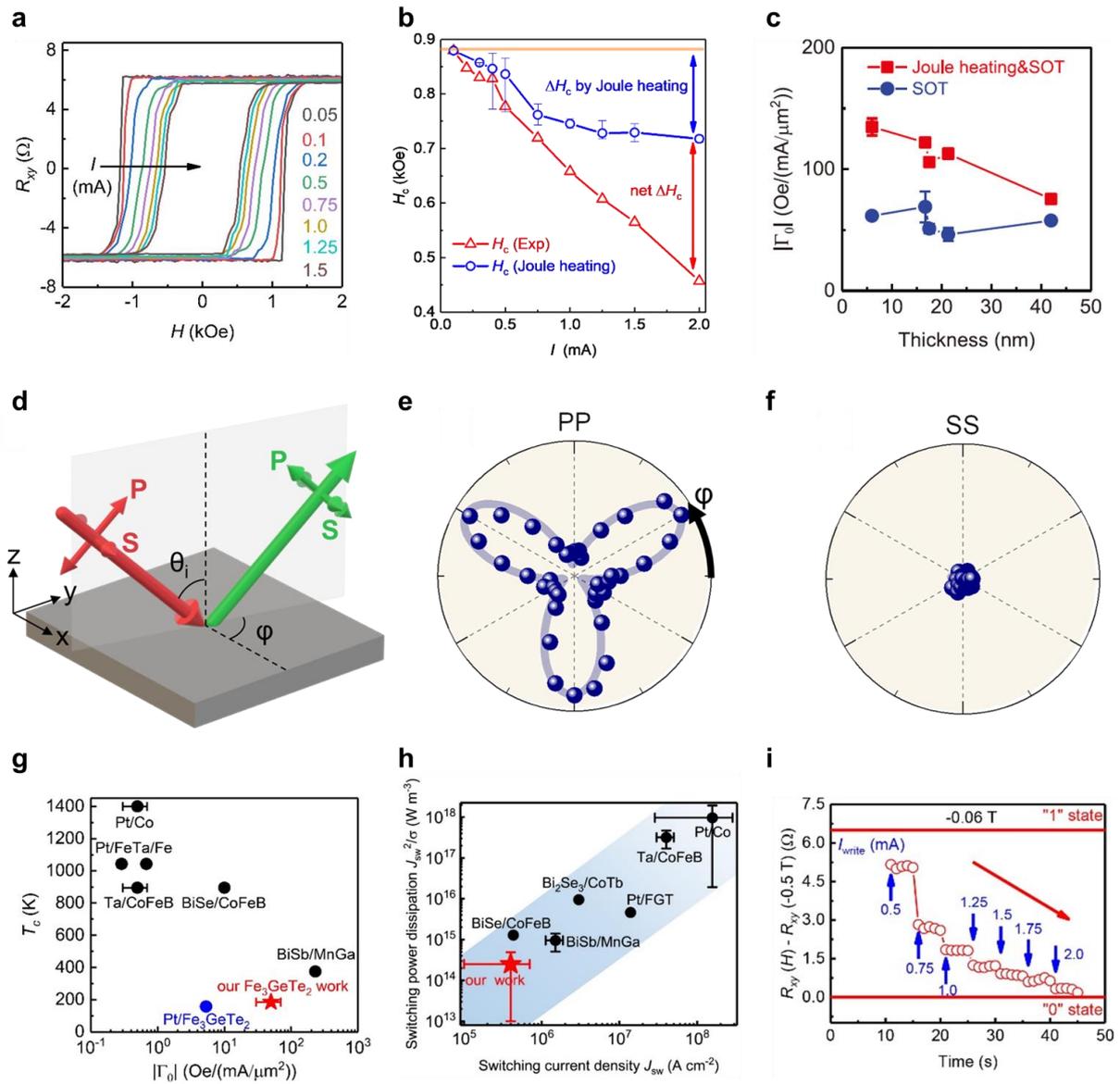

**Fig. 5. Gigantic intrinsic spin-orbit torque in $Fe_3GeTe_2$. a**, Increasing current induced substantial reduction of coercive field. **b**, Large coercivity reduction after Joule heating's contribution subtracted off. **c**, Spin-orbit torque (SOT) magnitude extracted from coercivity reduction for devices of varying thickness. The net SOT magnitude (blue points) shows no noticeable thickness dependence, with device thickness ranging from 6 to 40 nm. **d**, SHG schematic on bulk $Fe_{2.8}GeTe_2$. **e-f**, SHG response for the PP mode (**e**) shows a prominent three-fold petal, while for the SS mode (**f**) is nearly zero. It shows that inversion symmetry has been broken, mainly along the out-of-plane direction. **g**, Comparison of SOT magnitude of different systems, where the ME/SOT coefficient is about 100 times larger than that of conventional heavy metals. **h**, A novel type of magnetic memory has been demonstrated based on the gigantic intrinsic SOT. The switching current density and power dissipation are around 400 and 4000 times reduced compared to the conventional Pt/Co devices. **i**, Eight multi-level states controlled nonvolatilely by tiny currents, corresponding to 3 bits in a single-device magnetic memory.

(Fig. 5a-c is reproduced from Fig. 2a, Fig. S4f, and Fig. 3e respectively in Ref[17]; Fig. 5d-f is reproduced from Fig. 1a-c in Ref[19]; Fig. 5g is reproduced from Fig. 3f in Ref[17], and Fig. 5h,i is reproduced from Fig. 3d and Fig. 5c respectively in Ref[20].)

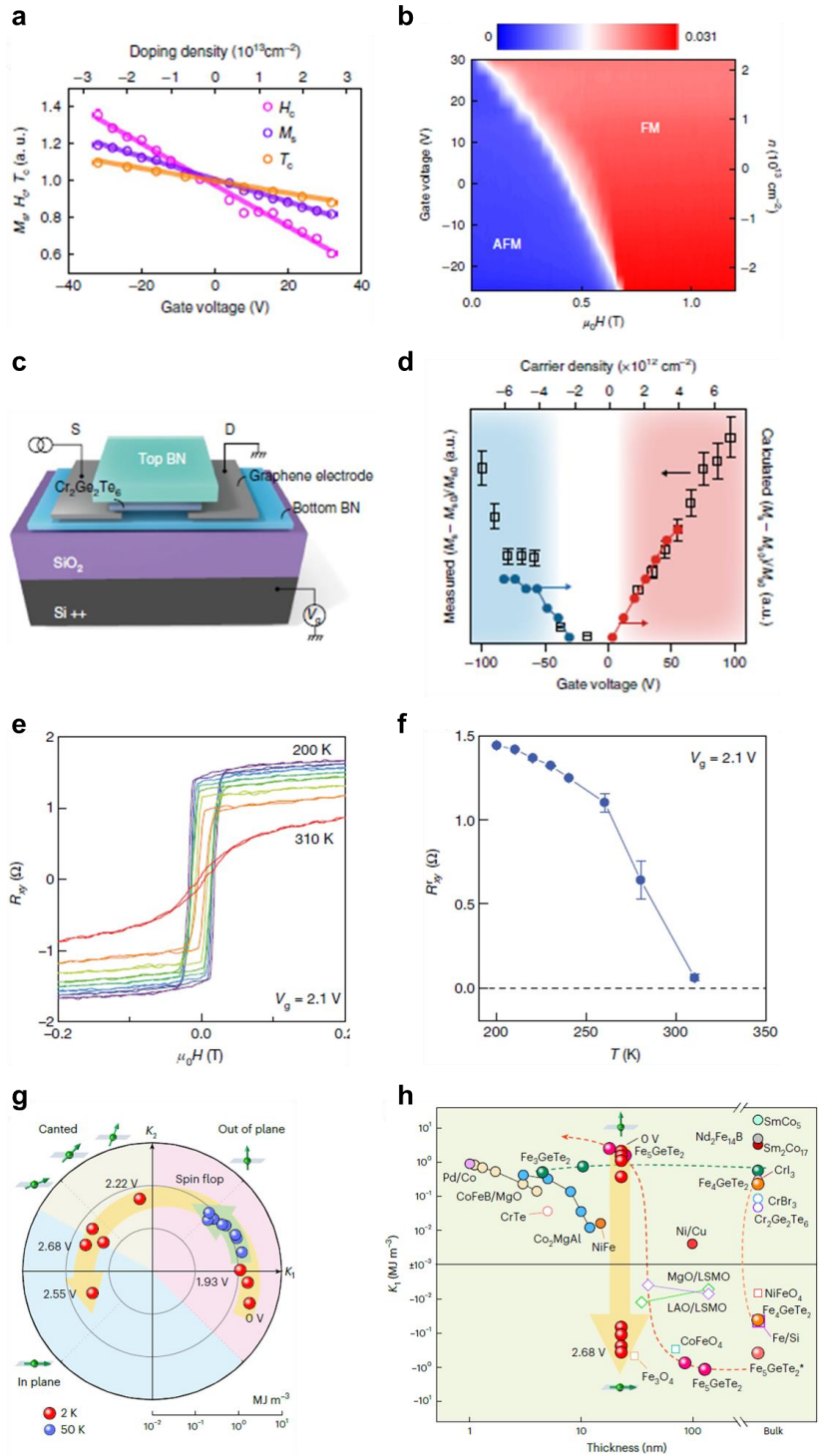

**Fig. 6. Four cases of gating vdW magnets. a**, Monolayer CrI$_3$'s saturated magnetization, coercivity, and Curie temperature can be linearly reduced by increasing gate voltage. **b**, Gating a bilayer CrI$_3$ leads to a antiferromagnetic to ferromagnetic transition. **c**, Back gate schematic for CrGeTe$_3$ nanoflakes. **d**, CrGeTe$_3$'s saturated magnetization can be enhanced by both positive and negative gate voltages. **e-f**, Ionic gating on a four-layer Fe$_3$GeTe$_2$, where the Curie temperature was boosted to 310 K at a gate voltage of 2.1 V. **g-h**, Ionic gating on a Fe$_5$GeTe$_2$

nanoflake tune the magnetic anisotropy from an out-of-plane to in-plane easy axis continuously. The $K_1$'s magnitude changes widely across a five-order range. (Fig. 6a,b is reproduced from Fig. 2c and Fig. 3b respectively in Ref[22]; Fig. 6c,d is reproduced from Fig. 3a and Fig. 5d respectively in Ref[24]; Fig. 6e,f is reproduced from Fig. 4a,b in Ref[25]; Fig. 6g,h is reproduced from Fig. 4a,b in Ref[26].)